\documentstyle[12pt]{article}
\newcommand{\be}{\begin{equation}}
\newcommand{\cE}{{\cal E}}
\newcommand{\ee}{\end{equation}}
\begin{document}
\begin{flushright}
DTP--MSU 98/01 \\ January 23, 1998
\end{flushright}
\vskip25mm\begin{center}
{\LARGE\bf
Generating branes via sigma-models
 }\\ \vskip1cm
{\bf
D.V. Gal'tsov}\footnote{email: galtsov@grg.phys.msu.su}
and
{\bf  O.A. Rytchkov}\footnote{email: rytchkov@grg1.phys.msu.su} \\
\normalsize  Department of Theoretical Physics,  Moscow State University,\\
\normalsize Moscow 119899, { Russia}\\
\vskip15mm
{\bf Abstract}\end{center}
\begin{quote}

Starting with the $D$-dimensional Einstein-dilaton-antisymmetric form 
equations and assuming a block-diagonal form of a metric we derive a  
$(D-d)$-dimensional $\sigma$-model with the target space
$SL(d,R)/SO(d) \times SL(2,R)/SO(2) \times R$ or its non-compact form. Various
solution--generating techniques are developed and applied to construct
some known and some  new $p$-brane solutions. It is shown that the Harrison 
transformation belonging to the $SL(2,R)$
subgroup generates black $p$-branes from the seed Schwarzschild
solution. A fluxbrane generalizing the Bonnor-Melvin-Gibbons-Maeda 
solution is constructed as well as a non--linear
superposition of the fluxbrane and a spherical black hole.
A new simple way to endow branes with additional internal
structure such as plane waves is suggested. Applying the harmonic maps
technique we generate new solutions with a non--trivial shell structure
in the transverse space (`matrioshka' $p$-branes). Similar $\sigma$-model
is constructed for the intersecting branes. It is shown that the
intersection rules have a simple geometric interpretation as
conditions ensuring the symmetric space property of the target
space. The null-geodesic method is used to find intersecting `matrioshka'
$p$-branes in Type IIA supergravity. Finally, a Bonnor-type symmetry relating
the four-dimensional vacuum $SL(2,R)$  with the corresponding sector of the
above global symmetry group is used to construct a new
magnetic 6-brane with a dipole moment in the ten-dimensional IIA theory.
\vskip5mm
\noindent
PASC number(s): 97.60.Lf, 04.60.+n, 11.17.+y
\end{quote}
\newpage
\section{Introduction}
Investigation of classical $p$-brane solutions to supergravities in
various dimensions has led to considerable progress in understanding
interlinks between different string models. Of particular interest
are IIA and IIB supergravities in ten dimensions which are the
low-energy limits of the corresponding superstring theories, and the
eleven-dimensional supergravity, which is supposed to be the low-energy limit
of M-theory. Recent progress in the string theory is also connected with
the discovery of non-perturbative objects called $D$- branes. In the
low-energy approximation they correspond to certain solutions
of appropriate classical equations.
There are several types of $p$-branes which
attracted attention. The  most important type includes
those purely bosonic solutions which preserve a part of the initial
supersymmetry, so called BPS states (for a review see \cite{Stelle,
Gaunt}). Another family consists of
not saturating Bogomol'nyi bound {\it black} $p$-branes possessing
a regular event horizon. Both families may form
intersecting multiple-brane structures. The solutions may be also endowed
with additional structures such as traveling waves.

Solving highly non-linear bosonic equations in the multidimensional
supergravities constitutes a formidable technical task.
In many cases $p$-brane solutions were obtained
using some special ans\"atze for the metric and matter fields
\cite{Ar_Vol, Iv1, ArRy, Englert}. In the BPS-saturated cases one can also
use the first order Bogomol'nyi type equations instead
of direct solving the equations of motion \cite{Duff1, Duff2}.
However this method is efficient mostly in eleven and ten-dimensional
supergravities where the Killing spinor equations are relatively simple.
Once such solutions are obtained, certain lower-dimensional solutions
may be found via appropriate compactification schemes. Some {\it ad hoc}
prescriptions are also known which allow to perform `blackening'
deformations of $p$-branes from extremal configurations
\cite{Duff_L_P, Cvetic_Ts, Ivanov}.
All these techniques are applicable only for
rather restricted classes of solutions.
\par
An approach which opens a way to explore more general solution classes
consists in the use of `hidden' symmetries (dualities) arising in
dimensionally reduced theories. This method allows to generate new
non--trivial solutions from known ones, as well as to suggest some new
integration algorithms. This approach appeared in the
four-dimensional General Relativity, where it has achieved
a high level of sophistication. For a class
of vacuum solutions effectively depending on three, rather than four
coordinates, the hidden symmetry group is $SL(2,R)$ which acts non-linearly
on two moduli. One of the symmetry
transformations (Ehlers transformation) is non--trivial and may be used as
generating symmetry. $N=2$ supergravity in four dimensions
being restricted to the class of solutions possessing a non--null Killing
vector field leads to the famous Kramer-Neugebauer-Kinnersley group
$SU(1,2)$ \cite{N_K, Kinner}, while the
(truncated to one vector) $N=4$ supergravity
generates the $Sp(4,R)$ symmetry \cite{Gal}. The crucial
role of three dimensions is due to the fact that the vector fields
can be traded there for scalars thus leading to a non--linear $\sigma$-model
description of the system.

The same approach can be used in higher dimensional supergravities
to construct multidimensional solutions. Here also the
$\sigma$-model description of the full space of solutions with a
sufficient number of commuting Killing vectors can be achieved
only in three dimensions. In higher dimensions
one has to deal with a great number of residual forms of various ranks,
originating from the initial forms and produced by the
Kaluza-Klein reduction. The symmetries of such reduced theories
are  called dualities ($U$-dualities).
It is well-known, for example, that the IIB supergravity, compactified to
nine dimensions, exhibits the $SL(2,R)$ symmetry ($S$-duality) mixing
the NS and RR fields. Also there is a correspondence between
the IIA and IIB supergravities reduced to  nine dimensions,
which is called $T$-duality \cite{Berg_Hull_Ort}. Using these
dualities, accompanied by appropriate boosts (or by the dimensional
reduction and uplifting), it is possible to construct a variety
of new solutions from the known ones \cite{R_Ts,Ts1,Ohta_Z}.
To fully exploit global symmetries arising due to dimensional reduction
in the $p$-brane context one has to construct explicit non--linear
realisations of the $U$-duality groups on the space of physical variables,
what is generally a highly non--trivial problem.
The first step towards this goal
is to consider a truncated
theory, in which only scalar fields are exited (what leads
to certain restrictions on the metric and the initial forms).
In this case we obtain a rather simple non-linear sigma-model, which
can be exhaustively analysed and fruitfully exploited.
\par
Starting with the $D$-dimensional Einstein-dilaton-antisymmetric form
equations and assuming a block-diagonal form of a metric we construct a
$\sigma$-model on the transverse space (of any dimension) with the isometry
(duality) group $SL(d,R) \times SL(2,R) \times R$.
Applying non--trivial transformations of this group
one can generate charged $p$--brane solutions from the seed
(multidimensional) Schwarzschild metric, to find a $p$-brane
generalization of the Melvin solution (a fluxbrane), to generate
intersecting branes and to put plane waves on branes.
Apart from a direct application of the target space isometries, one
can use $\sigma$-model approach to develop alternative integration
schemes, such as harmonic maps onto geodesic subspaces etc.
Ultimately one can find a completely integrable system
assuming dependence of solutions on only two variable.
\par
The paper is organized as follows. In Section 2 we consider the simple
containing $p$-branes bosonic theory, which describes the gravity coupled
$d$-form field and the dilaton. Using a block-diagonal ansatz
for the metric we
derive the corresponding $\sigma$-model action and examine its symmetries.
Section 3 is devoted to generation of the general black $p$-brane
solution by applying $\sigma$-model transformations. We argue that
the prescription of `blackening'  the extremal $p$-branes is a
manifestation of the target space isometries. In Section 4 using the
$\sigma$-model transformations we generate a fluxbrane, which is a
multidimensional analog of the Bonnor--Melvin universe. We also find
the non--linear superposition of the fluxbrane and a black hole.
In Section 5 we apply the technique of harmonic maps to obtain new
solutions of the $p$-brane type and study their properties. In Section
6 we discuss the {\it intersecting} $p$-branes in the $\sigma$-model terms.
It is shown that the well-known intersection rules restricting
dimensionalities and the coupling constants for known classes of
composite $p$-branes are equivalent to the symmetric space condition
for the target space. In this case the coset models may be formulated
which open a way to construct more general classes of intersecting
branes, an example is given for the case of the IIA supergravity.
In Section 7 we use the null geodesic method to generate
the Brinkmann wave and demonstrate its independence on the $p$-brane
structure. In Section 8 we discuss a Bonnor-type map relating
four-dimensional solutions of the vacuum Einstein equations to
multidimensional $p$-brane type solutions and derive an apparently
new  $p$-brane solution to the IIA supergravity in ten dimensions
endowed with a dipole moment. We conclude with some remarks on further
perspectives of the suggested approach. 

\section{Sigma-model representation}
Except for some particular applications to ten-dimensional IIA supergravity
in Sections 5 and 7 we will consider the model theory with the following
action  in the $D$-dimensional spacetime
\be
S=\frac{1}{2\kappa^2}\int\,d^Dx\sqrt{-G}\left(R-\frac12(\nabla\phi)^2-
\frac{{\rm e}^{-\alpha\phi}}{2(d+1)!}F^2_{d+1}\right),
\ee
where $F_{d+1}$ is a $(d+1)-$ differential form,
 $F_{d+1}=d{\cal A}_{d}$, $\phi$ is a dilaton. The corresponding
equations of motion are
\be
\label{Einstein}
R_{MN}-\frac12 G_{MN}R={\rm e}^{-\alpha\phi}T_{MN}^{(F)}+T_{MN}^{(\phi)},
\ee
\be
\label{Maxwell}
\partial_M\left({\rm e}^{-\alpha\phi}\sqrt{-G}F^{MM_1\ldots M_d}_{d+1}\right)=0,
\ee
\be
\label{dilaton}
\partial_M(\sqrt{-G}G^{MN}\partial_N\phi)+\frac{\alpha}{2(d+1)!}
{\rm e}^{-\alpha\phi}F^2_{d+1}=0.
\ee
The energy-momentum tensors for the matter fields have the form
\be
T_{MN}^{(F)}=\frac{1}{2d!}\left(F_{MM_1\ldots M_d}{F_N}^{M_1\ldots M_d}-
\frac{G_{MN}}{2(d+1)}F^2_{d+1}\right),
\ee
\be
T_{MN}^{(\phi)}=\frac12\left(\partial_M\phi\partial_N\phi-\frac12g_{MN}
(\nabla\phi)^2\right).
\ee
\par
Let us suppose that the space-time has $k$ commuting Killing vectors
orthogonal to hypersurfaces, one
of them being time-like (the case with  only space-like Killing
vectors will be discussed below). Then we can use the following
ansatz for the metric
\be
\label{metric}
 ds^2=g_{\mu\nu}(x)dy^\mu
dy^\nu+(\sqrt{-g})^{-\frac{2}{s}}h_{\alpha\beta}(x) dx^\alpha dx^\beta,
\ee
where $g_{\mu\nu}$  and $h_{\alpha\beta}$ are arbitrary $d$- and
$s+2$-dimensional metrics, with $\mu$, $\nu$ running from
$0$ to $d-1$, and $\alpha$, $\beta$ running from
$1$ to $s+2$, $s\ge 1,\; D=d+s+2$, $g=\det(g_{\mu\nu})$. The factor
$(\sqrt{-g})^{-\frac{2}{s}}$ is introduced for future convenience.
Both metric tensors depend only on (transverse) coordinates $x^\alpha$.

For the antisymmetric form we assume either electric or magnetic ans\"atze.
In the electric case the $d$-form has only one non-trivial
component
\be
\label{A}
A_{01\ldots d-1}=v(x).
\ee
With this choice of the metric and the $d$-form one obtains a
reduced theory in $s+2$-dimensional space. Clearly this reduction is not
the most general one, namely we have tacitly assumed that all Kaluza-Klein
vectors as well as the lower-dimensional
antisymmetric forms arising in full dimensional reduction are not excited.
However this truncated theory is still reach enough to be explored
in details.
\par
In terms of the functions $g_{\mu\nu}$, $h_{\alpha\beta}$, $\phi$ and
$v$ the equations of motion read
\be
\label{eq_g}
\frac{1}{\sqrt{h}}\partial_\alpha(\sqrt{h}h^{\alpha\beta}\partial_\beta
g_{\mu\lambda}g^{\lambda\sigma})g_{\sigma\nu}=\frac{s}{d+s}
{\rm e}^{-\psi}g_{\mu\nu}
h^{\alpha\beta}\partial_\alpha v\partial_\beta v,
\ee
\be
\label{eq_v}
\partial_\alpha(\sqrt{h}h^{\alpha\beta}{\rm e}^{-\psi}\partial_\beta v)=0, \ee
\be
\label{eq_phi}
\partial_\alpha(\sqrt{h}h^{\alpha\beta}\partial_\beta\phi)=\frac{\alpha}{2}
{\rm e}^{-\psi}h^{\alpha\beta}\partial_\alpha v\partial_\beta v,
\ee
$$
R_{\alpha\beta}^{(h)}=\frac{1}{2}\partial_\alpha\phi\partial_\beta\phi+
\frac{1}{s}\partial_\alpha(\ln\sqrt{-g})\partial_\beta(\ln\sqrt{-g})$$
\be
\label{eq_h}
+\frac14
g^{\mu\lambda}\partial_\alpha
g_{\lambda\nu}g^{\nu\sigma}\partial_\beta g_{\sigma\mu}-
\frac12{\rm e}^{-\psi}h^{\alpha\beta}\partial_\alpha v\partial_\beta v,
\ee
where
\be \label{Psi}
\psi=\alpha\phi+2\ln\sqrt{-g}.
\ee
It is straightforward to check that the field equations (\ref{eq_g}),
(\ref{eq_v}), (\ref{eq_phi}) and (\ref{eq_h}) can be obtained from
a new action of the $\sigma$-model type
$$
S=\frac{1}{2\kappa^2}\int\,d^{s+2}x\sqrt{h}\Bigg\{R^{(h)}-h^{\alpha\beta}
\Big(\frac{1}{2}\partial_\alpha\phi\partial_\beta\phi+
\frac{1}{s}\partial_\alpha(\ln\sqrt{-g})\partial_\beta(\ln\sqrt{-g})$$
\be
\label{gen_sigma_f}
+\frac14
g^{\mu\lambda}\partial_\alpha
g_{\lambda\nu}g^{\nu\sigma}\partial_\beta
g_{\sigma\mu}-\frac12{\rm e}^{-\psi}h^{\alpha\beta}\partial_\alpha v
\partial_\beta v\Big)\Bigg\}.
\ee

The similar action can be obtained assuming purely magnetic ansatz
for the $d$-form
\be
\label{F}
F^{\alpha_1\ldots\alpha_{s+1}}=\frac{1}{\sqrt{-G}}{\rm e}^{\alpha\phi}
\epsilon^{\alpha_1\ldots\alpha_{s+1}\beta}\partial_\beta u(x),
\ee
in this case one has to set in the metric $s=d$. The Maxwell
equations (\ref{Maxwell})  are trivially satisfied,
while the equation for $u$ follows from the Bianchi identity.
In this case the $\sigma$-model action still has the form
(\ref{gen_sigma}) with the replacement of $v$ on $u$ and reversing
the sign of $\alpha$. This fact is a manifestation of the electric-magnetic
duality. In what follows we consider explicitly an electric case,
the corresponding magnetic solutions can be obtained by the above
dualization procedure.

For subsequent analysis of the action (\ref{gen_sigma}) it is convenient
to renormalize the world-volume metric $g_{\mu\nu}$ introducing the matrix
\be
\label{tilde_g}
\tilde g_{\mu\nu}=(\sqrt{-g})^{-\frac2d} g_{\mu\nu},
\ee
such that $\det(\tilde g_{\mu\nu})=-1$.  Then the action will read
$$
S=\frac{1}{2\kappa^2}\int\,d^{s+2}x\sqrt{h}\Bigg\{R^{(h)}-h^{\alpha\beta}
\Big(\frac{1}{2}\partial_\alpha\phi\partial_\beta\phi+
\frac{s+d}{sd}\partial_\alpha(\ln\sqrt{-g})\partial_\beta(\ln\sqrt{-g})$$
\be
\label{gen_sigma}
+\frac14
\tilde g^{\mu\lambda}\partial_\alpha
\tilde g_{\lambda\nu}\tilde g^{\nu\sigma}\partial_\beta
\tilde g_{\sigma\mu}-\frac12{\rm e}^{-\psi}
\partial_\alpha v\partial_\beta v\Big)\Bigg\}.
\ee
Note, that the matrix $\tilde g_{\mu\nu}$ now is decoupled from the
rest of the $\sigma$-model variables, interacting with them only through
the gravitational field $h_{\alpha\beta}$. Since $\tilde g_{\mu\nu}$ is
a symmetric matrix with (minus) unit determinant
(the sign of the determinant is in fact irrelevant since the action remains
unchanged under a multiplication of $\tilde g_{\mu\nu}$ on a constant
matrix with the determinant minus one), this matrix parametrizes a coset
$SL(d,R)/SO(1,d-1)$. Therefore the metric on the world-volume of
the $p$-brane is to high extent independent of the other $\sigma$-model
variables, which only influence its determinant.  
\par 
To simplify the rest of
the action we introduce together with (\ref{Psi}) another variable
\be \label{xi}
 \xi=sd\phi-\alpha(s+d)\ln\sqrt{-g},
\ee
so that the inverse transformations read
\be \label{inv1}
\phi=\frac{1}{\Delta}\left(\alpha\psi+\frac{2\xi}{(s+d)}\right),
\ee
\be
\label{inv2}
\ln\sqrt{-g}=\frac{1}{\Delta (s+d)}\left(sd \psi-\alpha\xi\right),
\ee
where $\Delta=\alpha^2+2sd/(s+d)$.

In the new variables the part of the action not including the
matrix $\tilde g_{\mu\nu}$ will read
\be
\label{act_tr}
S=\frac{1}{2\kappa^2}\int\,d^{s+2}x\sqrt{h}\left\{R^{(h)}-h^{\alpha\beta}
\left(A\partial_\alpha\xi\partial_\beta\xi
+B\partial_\alpha\psi\partial_\beta\psi
-\frac12{\rm e}^{-\psi}\partial_\alpha v\partial_\beta v
\right)\right\},
\ee
where
\be
\label{AB}
A=\frac{1}{\alpha^2sd(s+d)+2s^2d^2},\qquad
B=\frac{s+d}{2\alpha^2(s+d)+4sd}.
\ee
Now the $\xi$-part is also decoupled. The remaining fields
$\psi$ and $v$ parametrize a coset $SL(2,R)/$$SO(1,1)$. Therefore
the action (\ref{gen_sigma}) corresponds to a non-linear $\sigma$-model
with the target space
$SL(d,R)/SO(1,d-1)$$\times SL(2,R)/SO(1,1)$$\times R$.
\par
Note that the possibility of description in the $\sigma$-model
terms on coset target spaces is typical for many dimensionally
reduced gravitational theories. The four-dimensional
Einstein--Maxwell theory  in the presence of one non-null Killing
vector field is equivalent to the $SU(2,1)/S(U(2)\times U(1))$
$\sigma$-model \cite{N_K, Kinner}. More complicated example is
given by the dilaton--axion coupled Einstein--Maxwell theory, in which
case one has the coset target space $Sp(4,R)/U(2)$ \cite{Gal}.
Several $\sigma$-models were derived in multidimensional
supergravities \cite{Iv1,Iv2,Iv3}, but the geometric structure of the target
spaces was not studied.
\par
Let us discuss  our  $\sigma$-model
(\ref{gen_sigma}) in  details. Since the potential space is
the direct product of three independent cosets, one can analyse
each of them separately. The transverse $SL(2,R)/SO(1,1)$ part
can be conveniently described by an analog of the Ernst potentials
\cite{Ernst}
\be \label{Phi}
\Phi=\frac{v}{2\sqrt{2B}},\qquad {\cal E}={\rm e}^{\psi}-\frac{v^2}{8B},
\ee
using which the target space metric can be rewritten in a familiar form
\cite{Kinner}
\be \label{act_E}
dl^2=\frac{1}{2}F^{-2}(d{\cal E}+2\Phi d\Phi)^2-2F^{-1}d\Phi d\Phi,
\ee \be F={\cal E}+\Phi^2.
\ee
The action of $SL(2,R)$ on the potentials is realized non-linearly.
It can be presented in terms of the following three one-parametric subgroups
\begin{eqnarray}
&\mbox{I.}&\; \cE=a^2\cE_0,\;\Phi=a\Phi_0,\\
&\mbox{II.}&\; \cE=\cE_0-2b\Phi_0-b^2,\;\Phi=\Phi_0+b,\\
&\mbox{III.}&\; \label{Harrison} {\cal
E}'=\frac{{\cal E}}{1-2c\Phi-c^2{\cal E}},\qquad
\Phi'=\frac{\Phi+c{\cal E}}{1-2c\Phi-c^2{\cal E}}, \end{eqnarray}
where $a,b$ and $c$ are parameters. Transformations I and II are
pure gauge ones, while the third (Harrison transformation) acts
on the space-time variables and matter fields non-trivially.
\par
Similarly one can consider the symmetry transformations
realized on the variables $\xi$  and $\tilde g$. Subgroup $R$ acts only on
 $\xi$:  $$\xi \rightarrow \xi+a.$$ In terms of initial fields it
corresponds to the shift of the dilaton on a constant accompanied by the
rescaling of the metric. The matrix $\tilde g$ parametrizes the coset
$SL(d,R)/SO(1,d-1)$, the representation of the group
$SL(d,R)$ is realized in a natural way
$$\tilde g\rightarrow U^{-1}\tilde g U,$$
where $U$ is a constant element of $SL(d,R)$.
\par
So far we have considered the equations of motion (\ref{Einstein}),
(\ref{Maxwell}), (\ref{dilaton}) assuming that the space-time metric
admits $d$ commuting Killing vectors one of which is time-like.
One can also investigate the case when all Killing vectors are
space-like.  In this case the $\sigma$-model action  will read $$
S=\frac{1}{2\kappa^2}\int\,d^{s+2}x\sqrt{-h}\Bigg\{R^{(h)}-h^{\alpha\beta}
\Big(A\partial_\alpha\xi\partial_\beta\xi
+\frac14
\tilde g^{\mu\lambda}\partial_\alpha
\tilde g_{\lambda\nu}\tilde g^{\nu\sigma}\partial_\beta
\tilde g_{\sigma\mu}$$
\be
+B\partial_\alpha\psi\partial_\beta\psi
+\frac12{\rm e}^{-\psi}\partial_\alpha v\partial_\beta v
\Big)\Bigg\},
\ee
where $A$ and  $B$ are the same (\ref{AB}).  This action differs from
the previous one by the sign of the last term. As a result the metric
on the target space is positively definite so we deal with the
coset $SL(d,R)/SO(d)\times SL(2,R)/SO(2)\times R$.
Introducing modified Ernst potentials
for the $SL(2,R)/SO(2)$ sector
\be \label{Phi'}
\Phi=\frac{v}{2\sqrt{2B}},\qquad {\cal E}=-{\rm e}^{\psi}-\frac{v^2}{8B},
\ee
we come back to the Eq. (\ref{act_E}).  Hence the Harrison transformation
again is given by (\ref{Harrison}).

\section{Generation of black $p$-branes}
As the first application of the above method we will consider
the  generation of black $p$-branes. Black $p$-brane is a multidimensional
generalization of the Reissner-Nordstr\"om black hole. The simplest
case $p=1$ corresponds to the black string \cite{Horowitz}. Another
important example is the black membrane of $D=11$ supergravity
\cite{Guven}. Black $p$-branes for general dimensions were constructed
in \cite{Duff_L_P,Duff_Lu}. Also there is a great number of papers
where the intersections of black $p$-branes are considered (see for
example \cite{Cvetic_Ts, Ivanov, Ohta} and references therein).
\par
Black $p$-branes are usually treated as a special deformation of the
corresponding extremal $p$-branes, specified by 
the one-center harmonic functions.
The process of deformation is called `blackening',  the relevant
prescription was gievn in \cite{Duff_L_P,Cvetic_Ts}.
Also it is known that black $p$-brane solutions can be obtained from
Schwarzschild solution by sequences of boosts and dualities \cite{Ts_Sch}.
We demonstrate
that the existence of such prescriptions is a manifestation of the hidden
symmetry contained in the model. This symmetry is nothing but
the $SL(2,R)$, which was considered in the previous section. In this
Section we use this symmetry for an explicit generation of a single
black $p$-brane, while in Section 6 we will explain the  generation
of intersecting black $p$-branes.
\par
Let us start with the Schwarzschild solution in the
$D$-dimensional spacetime correponding to a `neutral' $(d-1)$-brane 
\be
ds^2=-\left(1-\frac{2M}{r^s}\right)dt^2+dy_1^2+\ldots+dy_{d-1}^2+
\left(1-\frac{2M}{r^s}\right)^{-1}dr^2+r^2d\Omega_{s+1}.
\ee
Using the equations (\ref{Psi}) and (\ref{xi}) we obtain
\be
\psi_0=\ln\left(1-\frac{2M}{r^s}\right),\qquad \xi_0=-\frac12\alpha(s+d)
\ln\left(1-\frac{2M}{r^s}\right),\qquad v_0=0,
\ee
what coresponds to the following seed Ernst potentials
\be
\Phi_0=0,\qquad \cE_0=1-\frac{2M}{r^s}.
\ee
The Harrison transformations  (\ref{Harrison}) and the rescaling of
potentials yield the new functions $\psi$ and $v$
\be
\psi=\ln\left(1-\frac{2M}{r^s}\right)+
\ln\left(1+\frac{2Q}{r^s}\right)^{-2},\qquad
v=2c\sqrt{2B}\left(1-\frac{2M}{r^s}\right)
\left(1+\frac{2Q}{r^s}\right)^{-1},
\ee
where
\be
Q=\frac{Mc^2}{1-c^2}.
\ee
The function $\xi$ remains the same. The resulting metric is
$$
ds^2=\left(1+\frac{2Q}{r^s}\right)^{-\nu s}
\left\{-\left(1-\frac{2M}{r^s}\right)dt^2+
dy_1^2+\ldots+dy_{d-1}^2\right\}$$
\be
+\left(1+\frac{2Q}{r^s}\right)^{\nu d}\left\{
\left(1-\frac{2M}{r^s}\right)^{-1}dr^2+r^2d\Omega_{s+1}\right\},.
\ee
where $\nu=4\Delta^{-1}(s+d)^{-1}$.
It coincides with the metric of the black $p$-brane solution
\cite{Duff_L_P}. The corresponding dilaton field is given by
\be
{\rm e}^{-\alpha\phi}=\left(1+\frac{2Q}{r^s}\right)^{2\alpha^2/\Delta}.
\ee
Note that the extremal limit of this solution is $M\rightarrow 0$,
$c\rightarrow 1$ so that $Q$ is finite.

\section{Generation of the fluxbrane}
Assuming that all Killing vectors are space-like we can apply the same
technique to obtain the solution which is called a fluxbrane.
The fluxbrane is a multidimensional generalization of the Bonnor--Melvin
solution \cite{Melvin}, which is well-known in the usual Einstein-Maxwell
gravity. The Bonnor--Melvin solution with a dilaton was constructed by
Gibbons and Maeda \cite{Gibb_Ma}. We give its generalization
for the case of the arbitrary rank $d$-form of either electric,
or magnetic type.
\par
Our starting point is a flat $D$- dimensional space-time presented in the
multicylindrical coordinates
\be
ds^2=-dt^2+(\rho_1^2d\varphi_1^2+\ldots+\rho_d^2d\varphi_d^2)+d\rho_1^2+\ldots
+d\rho_d^2+dx_\alpha dx^\alpha,
\ee
where $\alpha=1,\ldots,s+1-d$.
This yields
\be
\psi_0=2\ln\rho_1\ldots\rho_d,\qquad \xi_0=-\alpha(s+d)\ln\rho_1\ldots\rho_d,
\qquad v_0=0,
\ee
so that the corresponding Ernst potentials are
\be
\cE_0=-\rho_1^2\ldots\rho_d^2,\qquad \Phi_0=0.
\ee
Applying the electric Harrison transformation (\ref{Harrison})
we obtain
\be
\psi=2\ln\rho_1\ldots\rho_d-2\ln(1+c^2\rho_1^2\ldots\rho_d^2),\qquad
v=-\frac{2c\sqrt{2B}\rho_1^2\ldots\rho_d^2}{1+c^2\rho_1^2\ldots\rho_d^2},
\ee
with $\xi$ remaining the same. As a result we get the following metric
$$
ds^2=\left(1+c^2\rho_1^2\ldots\rho_d^2\right)^{-\nu s}
\left(\rho_1^2d\varphi_1^2+\ldots+\rho_d^2d\varphi_d^2\right)
$$
\be
+\left(1+c^2\rho_1^2\ldots\rho_d^2\right)^{\nu d}
\left(-dt^2+d\rho_1^2+\ldots+d\rho_d^2+dx_\alpha dx^\alpha\right).
\ee
The corresponding dilaton field is
\be
{\rm e}^{-\alpha\phi}=\left(1+c^2\rho_1^2\ldots\rho_d^2\right)^{2\alpha^2/\Delta},
\ee
while the $d$-form potential has the non-vanishing component
\be
A_{\varphi_1\ldots\varphi_d}=
-\frac{2c\sqrt{2B}\rho_1^2\ldots\rho_d^2}{1+c^2\rho_1^2\ldots\rho_d^2},
\ee
where the coefficient $B$ is given by (\ref{AB}).
\par
Applying similar technique one can easily construct more complicated
solutions. As an example let us derive the
metric describing a six-dimensional dilatonic black hole in the
magnetic field of the 1-fluxbrane. Now we start not with the flat
space-time, but with the six-dimensional Schwarzschild solution
writing the metric on the 4-sphere in the form \cite{Myr_Per}:
 $$
ds^2=-\left(1-\frac{2M}{r^3}\right)dt^2+\left(1-\frac{2M}{r^3}\right)^{-1}dr^2$$
\begin{equation}
+r^2(d\theta^2+\cos^2\theta d\psi^2+\sin^2\theta
d\varphi_1^2+\cos^2\theta\sin^2\psi d\varphi_2^2 ).
\end{equation}
According to (\ref{Psi}) and  (\ref{xi}),
\begin{equation}
\psi_0=2\ln\left[\left(1-\frac{2M}{r^3}\right)^{-1}r^2\sin\theta\cos\theta\sin\psi
\right],
\end{equation}
$$
\xi_0=-4\alpha\ln\left[\left(1-\frac{2M}{r^3}\right)^{-1}r^2\sin\theta\cos\theta\sin\psi
\right],
$$
thus the seed Ernst potentials have the form
$$
{\cal
E}_0=-\left(1-\frac{2M}{r^3}\right)^{-2}r^4\sin^2\theta\cos^2\theta\sin^2\psi,
\qquad \Phi_0=0.
$$
Using Harrison transformations (\ref{Harrison})  we obtain the new
solution with the metric
$$
ds^2=\left\{1+c^2\left(1-\frac{2M}{r^3}\right)^{-2}r^4
\sin^2\theta\cos^2\theta\sin^2\psi\right\}^{\frac{2}{\alpha^2+2}}
\Bigg\{-\left(1-\frac{2M}{r^3}\right)dt^2 $$
$$+\left(1-\frac{2M}{r^3}\right)^{-1}dr^2+r^2 d\theta^2+r^2\cos^2\theta
d\psi^2\Bigg\}$$
$$+\left\{1+c^2\left(1-\frac{2M}{r^3}\right)^{-2}r^4\sin^2\theta\cos^2\theta
\sin^2\psi\right\}^{-\frac{2}{\alpha^2+2}}r^2(\sin^2\theta
d\varphi_1^2+\cos^2\theta\sin^2\psi d\varphi_2^2).
$$
The corresponding dilaton field is
$$
{\rm e}^{-\alpha\phi}=\left\{1+c^2\left(1-\frac{2M}{r^3}\right)^{-2}r^4\sin^2\theta\cos^2\theta
\sin^2\psi\right\}^{\frac{2\alpha^2}{\alpha^2+2}},
$$
while the non-vanishing component of the 2-form potential is given by
$$
A_{\varphi_1\varphi_2}=-\frac{2c}{\alpha^2+2}
\frac{\left(1-\frac{2M}{r^3}\right)^{-2}r^4
\sin^2\theta\cos^2\theta\sin^2\psi}{\left\{
1+c^2\left(1-\frac{2M}{r^3}\right)^{-2}r^4
\sin^2\theta\cos^2\theta\sin^2\psi\right\}}.
$$
It is easy to see that the obtained solution is indeed a non-linear
superposition of the black hole and the fluxbrane. The limit
$c\longrightarrow 0$ returns  us back to the Schwarzshild solution,
while putting $M=0$ we recover  the fluxbrane.

\section{Harmonic maps}
For further analysis we rewrite the $\sigma$-model action
(\ref{gen_sigma}) in the following matrix form
\be \label{act_Mg}
S=\frac{1}{2\kappa^2}\int\,d^{s+2}x\sqrt{h}\left\{R^{(h)}+h^{\alpha\beta}
\left(2B\mbox{Tr}\partial_\alpha M\partial_\beta
M^{-1}+\frac14 \mbox{Tr}\partial_\alpha \tilde
g\partial_\beta \tilde g^{-1}\right) \right\},
\ee
where the matrix $M$ is built from the fields $\Psi$, $v$  and  $\xi$
as follows
\be
\label{M}
M=\exp(-\frac{\psi}{2})\left(
\begin{array}{ccc}
2&\frac{v}{2\sqrt{2B}}&0\\
\frac{v}{2\sqrt{2B}}&-\frac12(\exp{\psi}-\frac{v^2}{8B})&0\\
0&0&\exp{\frac{\psi}{2}+\frac{\xi}{\sqrt{sd(s+d)}}}.\\
\end{array}
\right)
\ee
This representation is a convenient starting point for an application
of the harmonic maps technique. In particular we will be interested here in
constructing solutions corresponing to the null geodesics of the target 
space \cite{Clem,C_G}.

Consider the transverse part of the action (\ref{act_Mg})
\be \label{act_M}
S=\frac{1}{2\kappa^2}\int\,d^{s+2}x\sqrt{h}\left(R^{(h)}+2Bh^{\alpha\beta}
\mbox{Tr}\left(\partial_\alpha M\partial_\beta M^{-1}\right)\right),
\ee
where $M$ is an element of the appropriate coset space $G/H$. The
equations of motion read
 \be \label{eq_M}
\frac{1}{\sqrt{h}}\partial_\alpha(\sqrt{h}h^{\alpha\beta}M^{-1}
\partial_\beta M)=0,
\ee
\be
\label{eq_R}
R_{\alpha\beta}^{(h)}=-2B\mbox{Tr}(\partial_\alpha M \partial_\beta
M^{-1}).
\ee
It was noticed \cite{N_K}, that if the matrix $M$ depends on $x$-coordinates
through a single function, $M=M(\sigma(x))$, then $\sigma(x)$
can be chosen  to  be a harmonic function on the curved space 
with the metric $h$, i.e.
\be
\frac{1}{\sqrt{h}}\partial_\alpha(\sqrt{h}h^{\alpha\beta}
\partial_\beta \sigma)=0.
\ee
The equation (\ref{eq_M}) then reduces to the matrix equation
\be
\frac{d}{d\sigma}\left(M^{-1}\frac{dM}{d\sigma}\right)=0,
\ee
whose solution  can be expressed  in the exponential form
\be
\label{exp_M}
M=M_0{\rm e}^{K\sigma},
\ee
where $K$ belongs to the Lie algebra of the group $G$, and $M_0 \in G/H$.
Substituting this  into the Einstein equations (\ref{eq_R})
one gets
\be
R_{\alpha\beta}^{(h)}=2B\mbox{Tr}(K^2)\partial_\alpha \sigma
\partial_\beta \sigma.
\ee
It is clear that in the particular case $\mbox{Tr}(K^2)$=0 the metric $h$
is Ricci-flat (and hence can be chosen flat).
This is a constructive way to build null--geodesic
solutions to an arbitrary $\sigma$-model. Let us apply it to our
model with the target space $SL(2,R)/SO(1,1)\times R$. Here we are
interested in the asymptotically flat solutions, so we choose the
harmonic function $\sigma$ such that $\sigma(\infty)=0$.  According
to the above general scheme, we present $M$ in the form
(\ref{exp_M}), where $M_0$ is an element of the coset
$SL(2,R)/SO(1,1)\times R$ and the generator
$K$ belongs to the algebra
$sl(2,R)\times R$. $M_0$ has to be taken corresponding to an assumed
asymptotic behaviour, the Eq.(\ref{M}) gives
\be 
M_0=\mbox{diag}(2,-\frac12,1).
\ee
A convenient parametrization of the matrix $K$ is 
\be K=\left(\begin{array}{ccc}
a&c&0\\
d&-a&0\\
0&0&b\\
\end{array}\right).
\ee
One has to distinguish two different cases: $\det K\ne 0$  and
$\det K=0$.
\par
{\bf 1) Degenerate case: $\bf\mbox{det}\; K=0$}
\par
The general constraint $\mbox{Tr}K^2=0$ gives   $2(a^2+cd)+b^2=0$.
Together with the restriction $\det K=0$ this means $b=0,\;a^2+cd=0$.
In terms of the matrix $K$ this leads to $K^2=0$, so
the exponentiation is essentially different from that in
the non-degenerate case:
\be
{\rm e}^{K\sigma}=I+K\sigma.
\ee
Therefore for the matrix $M$ one obtains
\be
M=\left(\begin{array}{ccc}
2+2a\sigma&2c\sigma&0\\
-\frac12 d\sigma&-\frac12  (1-a\sigma)&0\\
0&0&1\\
\end{array}\right).
\ee
This matrix should be symmetric what gives an additional
constraint on the parameters, so the resulting matrix depends on a
single parameter $a$ 
\be 
M=\left(\begin{array}{ccc}
2+2a\sigma&a\sigma&0\\ a\sigma&-\frac12  (1-a\sigma)&0\\ 0&0&1\\
\end{array}\right).
\ee
Comparing this with the Eqs. (\ref{M}) and (\ref{Psi}) we get
\be
\psi=-2\ln(1+a\sigma),\qquad \xi=0,\qquad
v=2\sqrt{2B}\left(1-\frac{1}{1+a\sigma}\right).
\ee
Since $\sigma$ is an arbitrary harmonic function,
it is defined up to a scale parameter. Thus without loss of generality
one can put $a=1$.
The resulting metric is
\be
\label{metric_ex}
ds^2=\left(1+\sigma\right)^{-\nu s}
\left(-dt^2+dy_1^2+\ldots+dy_{d-1}^2\right)
+\left(1+\sigma\right)^{\nu d}\left(
dx_1^2+\ldots+dx_{s+2}^2\right).
\ee
This metric is nothing but  the usual
$p$-brane solution with the harmonic function $H=1+\sigma$ \cite{DGHR}.
The corresponding  dilaton field  is given by
\be
{\rm e}^{-\alpha\phi}=(1+\sigma)^{2\alpha^2/\Delta}.
\ee
This solution  saturates the Bogomol'nyi bound
\be
\label{m_ex}
M=\frac{\Omega_{s+1}}{2\kappa^2}8sBQ.
\ee
 
{\bf 2) Non-degenerate case: $\bf\mbox{det}\; K\ne 0$}
\par
Once again we have a constraint $\mbox{Tr}K^2=0$, what  implies
$2(a^2+cd)+b^2=0$.
Performing a direct exponentiation one obtains
\be
M=\left(\begin{array}{ccc}
2\cos\frac{b\sigma}{\sqrt{2}}+\frac{2a\sqrt{2}}{b}
\sin\frac{b\sigma}{\sqrt{2}}&\frac{2c\sqrt{2}}{b}
\sin\frac{b\sigma}{\sqrt{2}}&0\\
-\frac{d\sqrt{2}}{2b}\sin\frac{b\sigma}{\sqrt{2}}&
-\frac12\cos\frac{b\sigma}{\sqrt{2}}+\frac{a\sqrt{2}}{2b}
\sin\frac{b\sigma}{\sqrt{2}}&0\\
0&0&{\rm e}^{b\sigma}\\
\end{array}\right).
\ee
This matrix should be symmetric, so taking into account the
constraints on the coefficients we obtain
\be
M=\left(\begin{array}{ccc}
2\frac{\sin(\sigma+\varphi)}{\sin\varphi}&\frac{\sin\sigma}{\sin\varphi}&
0\\
\frac{\sin\sigma}{\sin\varphi}&
\frac{\sin(\sigma-\varphi)}{2\sin\varphi}&0\\
0&0&{\rm e}^{\sqrt{2}\sigma}\\
\end{array}\right),
\ee
where we put $b=\sqrt{2}$ because of the scaling freedom for
the harmonic function and denoted
\be
\sin\varphi=\frac{1}{\sqrt{a^2+1}}.
\ee
The Eqs. (\ref{M}) and (\ref{Psi}) yield
\be
\xi=\sigma\sqrt{2sd(s+d)},\quad \psi=-2
\ln\left[\frac{\sin(\sigma+\varphi)}{\sin\varphi}\right],\quad
v=\frac{2\sqrt{2B}\sin\sigma}
{\sin\left(\sigma+\varphi\right)},
\ee
Now it is easy to construct the whole metric
$$
ds^2=\left[\frac{\sin\left(\sigma+\varphi\right)}{\sin\varphi}\right]^{-\nu s}
{\rm e}^{-\sqrt{\frac{
s(s+d)}{2d}}\nu\alpha\sigma}
\left(-dt^2+dy_1^2+\ldots+dy_{d-1}^2\right)$$
\be
+\left[\frac{\sin\left(\sigma+\varphi\right)}{\sin\varphi}\right]^{\nu d}
{\rm e}^{\sqrt{\frac{d(s+d)}{2s}}\nu\alpha\sigma}
\left(
dx_1^2+\ldots+dx_{s+2}^2\right),
\ee
and the dilaton field
\be
{\rm e}^{-\alpha\phi}=\left[\frac{\sin\left(\sigma+\varphi\right)}{\sin\varphi}
\right]^{\frac{2\alpha^2}{\Delta}}
{\rm e}^{-\sqrt{\frac{sd(s+d)}{2}}\nu\alpha\sigma}.
\ee
\par
The structure of this solution is similar to that of the usual $p$-brane,
but the metric functions are essentially different (for the $0$-brane
case see \cite{C_G}). The full $r$-range solution
contains a sequence of compact singular transverse hypersurfaces
lying between the subsequent roots $r_k$ of the equation
\be
\sigma(r_k)+\varphi=\pi k,\qquad k=1,2,\ldots
\ee
and forming `matrioshka'-type structure in the transverse space. 
Curvature invariants diverge at $r_k$.
The outer solution is asymptotically flat, and for it
one can calculate  the ADM mass and the Page charge. 
It easy to check that the  Bogomol'nyi bound is saturated indeed 
(as could be expected since the solution corresponds to a null geodesic
in the target space) if
the parameter $\varphi$ satisfies the constraints
\be
\sin(\varphi+\chi)=\sqrt{\frac{sd}{2(s+d)}}, \qquad
\cos\chi=\alpha\sqrt{2B}.
\ee
\par
As the realistic example let us take the IIA supergravity,
whose bosonic action in the Einstein frame is given by
$$
S=\frac{1}{2\kappa^2}\int\,d^{10}x\sqrt{-G}\left(R-\frac12(\nabla\phi)^2-
\frac{{\rm e}^{-\phi}}{2\cdot 3!}F^2_{3}-\frac{{\rm e}^{\frac{3\phi}{2}}}{2\cdot 2!}
F^2_{2}\right.$$
\be\label{IIA}\left.
-\frac{{\rm e}^{\frac{
\phi}{2}}}{2\cdot 4!}F^{'2}_{4}\right)-\frac{1}{4\kappa^2}\int\;
F_4\wedge F_4\wedge A_2,
\ee
where
\be
F'_4=dA_3+A_1\wedge F_3.
\ee
We will consider the NS part of the action consisting of
the metric, dilaton and 2-form. The usual extremal 1-brane solution
corresponds to the elementary NS-string and has the form
\be
ds^2=H^{-\frac34}(-dt^2+dy^2)+H^{\frac14}(dx_1^2+\ldots+dx_{8}^2),
\ee
The `matrioshka' 1-brane line element reads
$$
ds^2=\left[\frac{\sin\left(\sigma+\varphi\right)}{\sin\varphi}\right]^
{-\frac34}{\rm e}^{-\frac{\sqrt{3}}{4}\sigma}
\left(-dt^2+dy^2\right) $$\be
+\left[\frac{\sin\left(\sigma+\varphi\right)}{\sin\varphi}\right]^
{\frac14}{\rm e}^{\frac{1}{4\sqrt{3}}\sigma}\left(
dx_1^2+\ldots+dx_{8}^2\right),
\ee
while the dilaton is
\be
{\rm e}^{-\phi}=\left[\frac{\sin\left(\sigma+
\varphi\right)}{\sin\varphi}\right]^{\frac{1}{2}}
{\rm e}^{-\frac{\sqrt{3}}
{2}\sigma}.
\ee

\section{Intersecting $p$-branes}
In order to describe within the same approach the intersecting $p$-branes
we have to change our basic
ansatz (\ref{metric}), (\ref{A}). Now assume that the $d$-form has
two nontrivial components
\be
A_{01\ldots q-1\,q\ldots q+r-1}=v_1(x),\qquad  A_{01\ldots
q-1\,q+r\ldots q+2r-1}=v_2(x),
\ee
where $d=r+q$.
A suitable parametrization for the metric is
$$
 ds^2=g_{\mu\nu}^{(0)}(x)dy^\mu
dy^\nu+g_{ij}^{(1)}(x)dz_1^i
dz_1^j+g_{ij}^{(2)}(x)dz_2^i
dz_2^j$$
\be
+(\sqrt{-g^{(0)}}\sqrt{g^{(1)}}\sqrt{g^{(2)}})
^{-\frac{2}{s}}h_{\alpha\beta}(x)dx^\alpha dx^\beta, 
\ee
where  $g_{\mu\nu}^{(i)}$  and
$h_{\alpha\beta}$ are arbitrary  symmetric tensors,
$i,j=1\ldots r$. As in the Section 2 we substitute this ans\"atze
into the equations of motion and
obtain the corresponding $\sigma$-model. Introducing as in
(\ref{tilde_g}) the "internal" metrics $\tilde g^{(i)}$
\be 
g_{\mu\nu}^{(0)}=(\sqrt{-g^{(0)}})^{\frac2q}\tilde
g_{\mu\nu}^{(0)},\qquad
g_{ij}^{(1)}=(\sqrt{g^{(1)}})^{\frac2r}\tilde g_{ij}^{(1)},\qquad
g_{ij}^{(2)}=(\sqrt{g^{(2)}})^{\frac2r}\tilde g_{ij}^{(2)},
\ee
it is easy to obtain the following. These renormalized metric tensors are
decoupled from the rest of the $\sigma$-model and we find them only
in the sector
$\frac14\mbox{Tr}\partial_\alpha\tilde g^{(i)}\partial_\beta\tilde
g^{{(i)}^{-1}}$.
The rest of the  $\sigma$-model action reads
$$
S=\frac{1}{2\kappa^2}\int\,d^{s+2}x\sqrt{h}\Bigg\{R^{(h)}-h^{\alpha\beta}
\Big(\frac{1}{2}\partial_\alpha\phi\partial_\beta\phi+
\frac{s+q}{sq}\partial_\alpha(\ln\sqrt{-g^{(0)}})\partial_\beta
(\ln\sqrt{-g^{(0)}})$$
$$
+\frac{s+r}{sr}\partial_\alpha(\ln\sqrt{g^{(1)}})\partial_\beta
(\ln\sqrt{g^{(1)}})+\frac{s+r}{sr}\partial_\alpha(\ln\sqrt{g^{(2)}})\partial_\beta
(\ln\sqrt{g^{(2)}})$$
$$
+\frac{2}{s}\partial_\alpha(\ln\sqrt{g^{(0)}})\partial_\beta
(\ln\sqrt{g^{(1)}})+\frac{2}{s}\partial_\alpha(\ln\sqrt{g^{(0)}})\partial_\beta
(\ln\sqrt{g^{(2)}})$$
$$+\frac{2}{s}\partial_\alpha(\ln\sqrt{g^{(1)}})\partial_\beta
(\ln\sqrt{g^{(2)}})
-\frac12{\rm e}^{-\alpha\phi-2\ln
\sqrt{-g^{(0)}}-2\ln
\sqrt{g^{(1)}}}\partial_\alpha v_1\partial_\beta
v_1$$
\be
\label{gen_sigma2}
-\frac12{\rm e}^{-\alpha\phi-2\ln \sqrt{-g^{(0)}}-2\ln
\sqrt{g^{(2)}}}\partial_\alpha
v_2\partial_\beta v_2\Big)\Bigg\}.  
\ee
Now the target space is six-dimensional, it is
parametrized by  $\phi$, $\ln\sqrt{-g^{(0)}}$,
$\ln\sqrt{g^{(1)}}$, $\ln\sqrt{g^{(2)}}$, $v_1$ and $v_2$.
\par
The explicit solutions known for the intersecting branes
were found only assuming certain conditions on the parameters
(intersection rules). It turns out that these conditions
correspond to the symmetric space property of the sigma-model target space.
Let us remind that the metric space is called symmetric, if
the Riemann tensor is covariantly constant, i.e.
\be \label{sym}
\nabla_a R_{bcde}=0.
\ee
Straightforward calculations yield the following. All non-zero components
of the five-index tensor  $\nabla_a R_{bcde}$ are proportional to
$$
\exp(-\alpha\phi-2\ln\sqrt{-g^{(0)}}-2\ln\sqrt{g^{(1)}})
\exp(-\alpha\phi-2\ln\sqrt{-g^{(0)}}-2\ln\sqrt{g^{(2)}})
\left(\frac{\alpha^2}{2}+\frac{qs-r^2}{q+2r+s}\right).
$$
This means that the target space is symmetric when the parameters
$s$, $q$, $r$ and $\alpha$
satisfy the following condition
\be
\frac{\alpha^2}{2}+\frac{qs-r^2}{q+2r+s}=0.
\ee  
This condition  can be rewritten as
\be\label{inter} 
\frac{\alpha^2}{2}+q-\frac{d}{D-2}=0,  
\ee
showing that we deal with the usual $p$-brane
intersection rule  \cite{ArRy}. Thus, the $\sigma$-model approach
gives a simple geometrical interpretation of the intersection rule
(\ref{inter}): only when  (\ref{inter}) is satisfied, the
target space is a symmetric (pseudo)Riemannian space.
\par
If the parameters of our configuration satisfy (\ref{inter}),
the  $\sigma$-model
action (\ref{gen_sigma2})
could be diagonalized and reduced to the simple form similar
to (\ref{act_tr}):
$$
S=\frac{1}{2\kappa^2}\int\,d^{s+2}x\sqrt{h}\left\{R^{(h)}-h^{\alpha\beta}
\Big(A_1\partial_\alpha\xi_1\partial_\beta\xi_1+
A_2\partial_\alpha\xi_2\partial_\beta\xi_2
+B_1\partial_\alpha\psi_1\partial_\beta\psi_1\right.
$$
\be\label{act_tr2}\left.
-\frac12{\rm e}^{-\psi_1}\partial_\alpha v_1\partial_\beta v_1
+B_2\partial_\alpha\psi_2\partial_\beta\psi_2
-\frac12{\rm e}^{-\psi_2}\partial_\alpha v_2\partial_\beta v_2
\Big)\right\},
\ee
where
\be
\psi_1=\alpha\phi+2\ln\sqrt{-g^{(0)}}+2\ln\sqrt{g^{(1)}},
\ee
\be
\psi_2=\alpha\phi+2\ln\sqrt{-g^{(0)}}+2\ln\sqrt{g^{(2)}},
\ee
\be
\xi_1=-\alpha s\phi -2(s-r)\ln\sqrt{-g^{(0)}}+2r\ln\sqrt{g^{(1)}}
+2r\ln\sqrt{g^{(2)}},
\ee
\be
\xi_2=q(r+s)\phi-\alpha(q+2r+s)\ln\sqrt{-g^{(0)}},
\ee
and the constants are
\be
A_1=\frac{1}{4sr^2},\qquad A_2=\frac{1}{2(q+2r+s)r^2d},\qquad
B_1=\frac{1}{4r},\qquad B_2=\frac{1}{4r}.
\ee
Thus we have obtained the $\sigma$-model with the $SL(2,R)/SO(1,1)\times
SL(2,R)/SO(1,1)\times R\times R$ target space. This structure
means that two $p$-branes can be generated separately. As an example one
can construct two intersecting black $p$-branes. The procedure is similar
to that discussed in Section 3, but now one has to apply
Harrison transformations with different parameters to each $SL(2,R)/SO(1,1)$
component. Thus one obtains two intersecting non-extremal $p$-branes with
different charges \cite{Cvetic_Ts, Ivanov}. This derivation
demonstrates that the existence of such configurations is a consequence
of the $\sigma$-model target space symmetries.
\par
In the Sec. 5 we have constructed some new solutions using the null geodesic
method applied to the  $\sigma$-model (\ref{act_tr}). The same
strategy applied to the $\sigma$-model  (\ref{act_tr2}) leads to
extremal intersecting $p$-branes with two charges (in the case of the 
degenerate matrix $K$) and to the intersecting `matrioshka'-type $p$-branes
(in the case of the non-degenerate matrix $K$). Thus we can speculate
that `matrioshka' $p$-branes are subject of the usual intersection rule.
As an example we exhibit the metric of two intersecting `matrioshka'-type
1-branes in the Type IIA supergravity:
$$
ds^2=\left[\frac{\sin\left(\sigma_1+\varphi_1\right)}{\sin\varphi_1}\right]^
{\frac38}\left[\frac{\sin\left(\sigma_2+\varphi_2\right)}{\sin\varphi_2}\right]^
{\frac38}{\rm e}^{-\frac{7\sqrt{6}}{48}\sigma_1-\frac{\sqrt{2}}{16}\sigma_2}$$
$$\left(-\left[\frac{\sin\left(\sigma_1+\varphi_1\right)}{\sin\varphi_1}\right]^
{-1}\left[\frac{\sin\left(\sigma_2+\varphi_2\right)}{\sin\varphi_2}\right]^
{-1}{\rm e}^{-\frac{\sqrt{6}}{6}\sigma_1+\frac{\sqrt{2}}{2}\sigma_2}
\left(-dt^2+dy^2\right)\right. $$
\be\left.
+\left[\frac{\sin\left(\sigma_1+\varphi_1\right)}{\sin\varphi_1}\right]^
{-1}{\rm e}^{\frac{\sqrt{6}}{3}\sigma_1}\left(
dz_1^2+dz_{2}^2\right)+\left[\frac{\sin\left(\sigma_2+\varphi_2\right)}
{\sin\varphi_2}\right]^
{-1}{\rm e}^{\frac{\sqrt{6}}{3}\sigma_1}\left(
dz_3^2+dz_{4}^2\right)+dx_\alpha dx^\alpha\right).
\ee
\section{Brinkmann waves on branes}
\par
The null geodesic method can also be applied to the $SL(d,R)/SO(1,d-1)$ part
of the initial $\sigma$-model (\ref{gen_sigma}). The matrix $\tilde g$
should be taken in the form, similar to (\ref{exp_M})
\be
\tilde g=\tilde g_0 {\rm e}^{K\sigma},
\ee
where
$\mbox{Tr} K^2=0$, $K$ belongs to an algebra $sl(d,R)$. Asymptotic
flatness conditions imply
\be
\tilde g_0 = \mbox{diag} (-1, 1,\ldots).
\ee
In the simplest case $d=2$ the condition $\mbox{Tr} K^2=0$ leads to
$\det K=0$, i.e. the resulting matrix $\tilde g$ has the form
\be
\tilde g=\left(\begin{array}{cc}
-(1+a\sigma)&-c\sigma\\
d\sigma&1-a\sigma\\
\end{array}\right),\qquad a^2+dc=0.
\ee
The matrix $\tilde g$ should be symmetric, so $c=-d=\pm a$,
and after rescaling of the harmonic function $\sigma$
\be
\tilde g=\left(\begin{array}{cc}
-(1+\sigma)&\pm\sigma\\
\pm\sigma&1-\sigma\\
\end{array}\right).
\ee
This metric can be rewritten in the light-cone coordinates as
\be
ds^2=-du\,dv-\sigma du^2.
\ee
It corresponds to the well-known Brinkmann wave \cite{Brink}
and the decoupling of
$\tilde g$ from the action (\ref{gen_sigma}) reflects the possibility
of a superposition of $p$-branes and waves \cite{Berg}.

\section{Bonnor-type map}
One can obtain new non-trivial solutions from the old one
using a map between similar cosets  describing
physically different theories. This idea traces back to the
Bonnor construction of the metric of a magnetic dipole
in General Relativity using a correspondence
of two $SL(2,R)/SO(1,1)$ describing stationary vacuum gravity
and static electrovacuum. Since we have the same subspace
in the $p$-brane case (\ref{gen_sigma}), one can use the same
correspondence to generate new $p$-brane solutions.
 \par
For the vacuum Einstein theory in four dimensions the
target space describing stationary solutions has the form
\be
ds^2=\frac{1}{2f^2}(df^2+d\chi^2),
\ee
where $f=g_{tt},$ and $\chi$ is the twist-potential.
One can check that the
correspondence between two $\sigma$-models can be achieved only
if $B=1/8$. Note that the appropriate map is complex
\be
\label{comp_tr}
\Psi=2\ln f,\qquad v=i\chi,
\ee
so, in order to obtain real solutions in the Minkowskian space,
we should take the  complexified seed solutions.
\par
As an example let us consider the complexified  Kerr-NUT solution of
the Einstein theory taking pure imaginary rotation and NUT parameters
$\tilde a=ia$, $\tilde N=iN$
\be
ds^2=-\frac{\Delta+\tilde a^2\sin^2\theta}{\Sigma}(dt-\omega
d\varphi^2)+\Sigma
\left(\frac{dr^2}{\Delta}+d\theta^2+\frac{\Delta\sin^2\theta}
{\Delta+\tilde a^2\sin^2\theta}d\phi^2\right),
\ee
where
\be
\label{Delta}
\Delta=r^2-2Mr-\tilde a^2+\tilde N^2,
\ee
\be
\label{Sigma}
\Sigma=r^2+\delta^2,\qquad \delta=-i(\tilde a\cos\theta+\tilde N),
\ee
\be
\omega=\frac{2i}{\Delta+\tilde a^2\sin^2\theta}(\tilde N\Delta\cos\theta
+\tilde a\sin^2\theta(Mr-\tilde N^2)).
\ee
The potentials $f$ and $\chi$ will be given by
\be
\label{f_chi}
f=\frac{\Delta+\tilde a^2\sin^2\theta}{\Sigma},\qquad
\chi=-\frac{2i(M\tilde a\cos\theta+M\tilde N-\tilde Nr)}{\Sigma}.
\ee
Since the potential $\chi$ is pure imaginary, the complex
transformation  (\ref{comp_tr}) will give a real solution.
\par
Consider the Type IIA supergravity (\ref{IIA}).
It  contains a 1-form which can be connected with electric black
hole or with the magnetic D6-brane. We will construct the metric of
the D6-brane. It is easy to check that in this case $B$ is equal to 1/8,
so we can use the above technique. The map  (\ref{comp_tr})
applied to the potentials  (\ref{f_chi}) and the Eqs.
(\ref{inv1}), (\ref{inv2})  lead the following metric
\be
ds^2=f^{\frac18}(-dt^2+dy_1^2+\ldots+dy_6^2)+
f^{\frac18}\frac{\Sigma}{\Delta}dr^2+f^{\frac18}\Sigma d\theta^2
+f^{-\frac78}\Delta\sin^2\theta d\varphi^2,
\ee
where $f$, $\Delta$ and $\Sigma$ are given by (\ref{f_chi}), (\ref{Delta})
and (\ref{Sigma}).
This metric describes the magnetic D6-brane with the dipole moment
which is generated by the parameter $\tilde a$.
The non-trivial components  of the 1-form field strength  are
\be
F^{r\varphi}=-\frac{f^{-5/4}}{\Sigma\sin\theta}\partial_\theta u,\qquad
F^{\theta\varphi}=\frac{f^{-5/4}}{\Sigma\sin\theta}\partial_r u,
\ee
where  $u$ is given by
\be
u=\frac{2(M\tilde a\cos\theta+M\tilde N-\tilde Nr)}{\Sigma}.
\ee
The corresponding dilaton field can be expressed through the
function $f$ as follows
\be
{\rm e}^{\frac32\phi}=f^{\frac98}.
\ee
It is easy to check that  the Bogomol'nyi bound is saturated if
$M=N$. In the limit $\tilde a=0$, $M=N$  the  configuration obtained
reduces to the usual extremal magnetic D6-brane.

\section{Concluding remarks}
In this paper, we have focused  on the technical aspects of getting
$p$-brane solutions via the $\sigma$-model formulation of the
simplest brane-containing theories.
Although the idea of using dualities of dimensionally reduced
theories is not new, we have shown that knowing an explicit
non-linear realization of dualities in terms of the target space variables 
one can exploit `hidden symmetries' to higher extent. Apart from a direct
use of transformations to get new solutions from old ones, one can
also apply various integration methods developed earlier in
the context of General Relativity. In particular,
using a technique of harmonic maps we have found  new classes
of $p$-branes with a nontrivial `matrioshka'-type structure
of the transverse space. We have shown that some $p$-brane
`rules', such as intersection rules for composite branes, or
`blackening' prescriptions, have a rather natural geometric
interpretation in the $\sigma$-model terms. Since the main
subgroup involved is $SL(2, R)$, one can effectively use
solutions to other theories sharing the same group structure
to get new $p$-brane solutions, this Bonnor-type correspondence
is somewhat similar to duality between different theories which
was widely discussed recently in the context of the superstrings.

Our formulation also opens a way to apply techniques of integrable
systems assuming that the target space variables depend only
on two of the transverse coordinates. In the four-dimensional theories
the full space-time metric can be recovered once the solution
of the corresponding integrable system is found. In the multidimensional
cases additional assumptions are needed about the structure of the
transverse space to ensure complete solvability.
More general lagrangians including several antisymmetric forms
and dilatons may be also investigated under the assumption of
the block-diagonal metrics. However in the non--diagonal
cases one encounters serious technical complications while
attempting to find an explicit non--linear realization of
`hidden' symmetries.

\vskip1cm
{\bf Acknowledgements}
\bigskip

This work was supported in part by the Russian Foundation
for Basic Research grants 96-01-00608 (O.R.) and 96-02-18899 (D.G.).


\begin{thebibliography}{99}
\bibitem{Stelle} K.~S.~Stelle, {\it Lectures on Supergravity $p$-branes},
Report Imperial/TP/96-97/15, hep-th/9701088.
\bibitem{Gaunt} J.~P.~Gauntlett, {\it Intersecting Branes},
Report QMW-PH-97-13, NI-97023, hep-th/9705011.
\bibitem{Ar_Vol} I.~Ya.~Aref'eva and A.~I.~Volovich,
Class. Quant. Gravity {\bf14}, 1 (1997).
\bibitem{Iv1} V.~D.~Ivashchuk and V.~N.~Melnikov,
Grav. and Cosmol. {\bf 2}, 297 (1996).
\bibitem{ArRy} I.~Ya.~Aref'eva and O.~A.~Rytchkov,
Report SMI-25-96, hep-th/9612236.
\bibitem{Englert} R.~Arguirio, F.~Englert, and L.~Houart,
             Phys.Lett. {\bf B398}, 61-68 (1997).
\bibitem{Duff1} M.~J.~Duff, R.~R.~Khuri, and J.~X.~Lu,
Phys.Rep. {\bf 259}, 213 (1995).
\bibitem{Duff2} M.~J.~Duff, {\it Supermembranes}, TASI
96 lectures, Report CTP-TAMU-61/96, hep-th/9611203.
\bibitem{Duff_L_P} M.~J.~Duff, H.~L\"u and C.~N.~Pope,
Phys. Lett {\bf B382}, 73 (1996).
\bibitem{Cvetic_Ts} M.~Cveti\v{c} and A.~A.~Tseytlin,
Nucl. Phys. {\bf B478}, 181 (1996).
\bibitem{Ivanov} I.~Ya. Aref'eva, M.~G.~Ivanov, I.~V.~Volovich,
Phys. Lett. {\bf B406}, 44 (1997)
\bibitem{N_K} G.~Neugebauer and D. Kramer, Ann. der Physik (Leipzig)
{\bf 24}, 62 (1969).
\bibitem{Kinner} W.~Kinnersley, J. Math. Phys. {\bf 18}, 1529 (1977).
\bibitem{Gal} D.~V.~Gal'tsov, Phys. Rev. Lett. {\bf 74}, 2863 (1995) .
\bibitem{Berg_Hull_Ort} E. Bergshoeff, C. Hull and Ortin, Nucl. Phys.
{\bf B451}, 547 (1995)
hep-th/9504081
\bibitem{R_Ts} J.~G.~Russo and A.~A.~Tseytlin, Nucl. Phys. {\bf B490},
121 (1997)
\bibitem{Ts1} A.~A.~Tseytlin, Class. Quant. Grav. {\bf 14}, 2085
(1997)
\bibitem{Ohta_Z} N.~Ohta and J.~Zhou, {\it Towards the
Classification of Non-Marginal Bound States of M-branes and Their
Construction Rules}, Report OU-HET 267, hep-th/9706153.
  \bibitem{Iv2}  V.~D.~Ivashchuk
and V.~N.~Melnikov, Class. Quant. Grav. {\bf 14}, 3001 (1997)
 \bibitem{Iv3}
V.~D.~Ivashchuk, V.~N.~Melnikov, and M.~Rainer, {\it Multidimensional
$\sigma$-models with Composite Electric $p$-branes},
Report Uni-P-Math/4-5-97, gr-qc/9705005.
\bibitem{Ernst} F.~J.~Ernst, Phys. Rev. {\bf 167}, 1175 (1968);
{\bf 168}, 1415 (1968).
\bibitem{Horowitz} G.~T.~Horowitz and A.~Strominger,
Nucl. Phys. {\bf B360}, 197 (1991).
\bibitem{Guven} R.~Guven, Phys. Lett. {\bf B276}, 49 (1992).
\bibitem{Duff_Lu} M.~J.~Duff and J.~X.~Lu,
Nucl. Phys. {\bf B416}, 301 (1994).
\bibitem{Ohta} N.~Ohta, Phys. Lett. {\bf B403}, 218 (1997)
\bibitem{Ts_Sch} A. A. Tseytlin, Phys. Lett. {\bf B395}, 24 (1997)
\bibitem{Melvin} M.~A.~Melvin, Phys. Lett. {\bf 8}, 65 (1964).
\bibitem{Gibb_Ma} G.~W.~Gibbons and K.~Maeda, Nucl. Phys. {\bf B298}, 741
(1988).
\bibitem{Myr_Per} R.~C.~Myers and M.~Perry, Ann. Phys. {\bf 172}, 304 (1986).
\bibitem{Clem} G.~Clement, Gen. Rel. and Grav. {\bf 18}, 861 (1986);
Phys. Lett. {\bf A118}, 11 (1986).
\bibitem{C_G} G.~Clement and D.~Gal'tsov, Phys. Rev. {\bf 54}, 6136 (1996).
\bibitem{DGHR} A. Dabholkar, G.~W.~Gibbons, J.~A.~Harvey, and F.~Ruiz Ruiz,
Nucl. Phys. {\bf B340}, 33 (1990).
\bibitem{Lu} J.~X.~Lu, Phys. Lett. {\bf B313}, 29 (1993)
\bibitem{Brink} H.~W.~Brinkmann, Proc. Nat. Acad. Sci. {\bf 9}, 1 (1923).
\bibitem{Berg} E.~Bergshoeff, M.~de Roo, E.~Eyras, B.~Janssen, and
 J.~P.~van der Schaar, Class. Quant. Grav. {\bf 14}, 2757 (1997)

\end{thebibliography}
\end{document}